\documentclass{llncs}
\usepackage{subfig}
\usepackage{cite}
\usepackage{mathptmx}       
\usepackage{helvet}         
\usepackage{courier}        
\usepackage{type1cm}        
\usepackage{makeidx}         
\usepackage{graphicx}        
\usepackage{multicol}        
\usepackage[bottom]{footmisc}
\usepackage{amsfonts}
\usepackage[cmex10]{amsmath}
\begin{document}

\title{Sockpuppet Detection: a Telegram case study} 
\titlerunning{Sockpuppet detection: a Telegram case study}  
%

\author{Gabriele Pisciotta\inst{1}  \and Miriana Somenzi\inst{1} \and Elisa Barisani\inst{1} \and Giulio Rossetti\inst{4}}
\authorrunning{Gabriele Pisciotta et al.} 
%
\tocauthor{Gabriele Pisciotta, Miriana Somenzi, Elisa Barisani and Giulio Rossetti}
\institute{University of Pisa, Italy,\\
	\email{\{g.pisciotta1, m.somenzi, e.barisani\}@studenti.unipi.it}
	\and
	KDD Lab, ISTI-CNR, Italy \\
	\email{giulio.rossetti@isti.cnr.it}}

\maketitle

\section{Introduction}
In Online Social Networks (OSN) numerous are the cases in which users create multiple accounts that publicly seem to belong to different people but are actually fake identities of the same person. These fictitious characters can be exploited to carry out abusive behaviors such as manipulating opinions, spreading fake news and disturbing other users \cite{DBLP:journals/corr/SolorioHM13,sckpgraph}. 
In literature this problem is known as the "Sockpuppet problem". \\ 
In our work we focus on Telegram, a wide-spread instant messaging application, often known for its exploitation by members of organized crime and terrorism, and more in general for its high presence of people who have offensive behaviors. In Italy, for example, it stepped into the spotlight because of a revenge porn case in early 2020. In this OSN users can chat both privately and in groups, and its peculiarity is that they can interact anonymously, because the service allows you to display only your nickname of choice, keeping any other personal information hidden. This feature facilitates the emergence of various fake accounts, exploiting also free VoIP numbers to create a variety of sockpuppet accounts. \\
The detection of sockpuppet accounts is a challenging task and the approaches\cite{YAMAK2018124} that have been tried during these years involve many different platforms and OSN; furthermore these techniques can be tailored or applied in a more general way. The majority of them often relies on the use of Natural Language Processing (NLP) methods, OSN-dependent features and mined behavioral patterns. \\  Thus we decided to tackle the problem in an innovative way to try and have more interesting results, that is by involving Network Science as a tool to match different virtual users that are actually the same person offline. To do so we took several Italian public groups on Telegram, choosing them because of the similarity in their academic contents and the fact that these groups have some users in common (they actually belong to the same virtual network of self-declared "friends group"). We scraped these groups' messages and exploited the replies between users, all this in a completely undetectable way from the inside thanks to the fact that it is not necessary to join the groups to retrieve the informations. \\We then created a directed weighted network of user interactions that connects the users who write a message, the source, to the users to whom they reply, the target. We took 17747 users as nodes and 191526 explicit replies to messages as edges, weighted by the number of replies. 

\section{Preliminar Results}

We decided to investigate the power of Network Science applied on Sockpuppet Detection in Telegram. More in depth, we saw this problem as an instance of the link prediction task based on similarity measures, where the link has the meaning of "same-as". In agreement to this the whole problem can be seen as an instance of the data linking problem that, according to the particular scenario taken into account, can be named as deduplication (w.r.t. databases), instance matching (w.r.t knowledge graphs) and entity resolution: we decided to to treat it in a similar way, trying to find cluster of similar instances. \\
We propose a scalable method that exploits neighbours as features. In order to assess the similarity between the nodes we should compute the similarity matrix according to a certain function and filter that need to be the closest possible to a given threshold. However this operation is very costly, both in memory and time, being O($n^2$) (where n is the number of nodes).  To tackle this scalability issue we involved a family of algorithms known as Locality-Sensitive Hashing (LSH)\cite{10.1145/276698.276876} that create blocks in which similar entities are stored based on an hash function that takes each entity's feature as input: similar entities should lead to similar hash, so they should belong to the same cluster. Instead of computing the similarity between each pair of nodes, using LSH algorithms we drop from a complexity of O($n^2$) to O(n), allowing us to scale w.r.t the number of nodes. In specific we use \texttt{SimHash} \cite{10.1145/1242572.1242592}, to approximate the hamming distance between the the nodes and get the clusters. \\
Before computing the similarity between nodes, we preprocessed the data in this way:
\begin{itemize}
    \item we normalized neighbours' weight according to each user
    \item we dropped links having weight less than 0.5 to only take into account the more meaningful interactions
\end{itemize}
For each node we wanted to limit the number of similar ones to have few candidates: to do this, we set the fingerprint size to 128 and the max distance to 20.\\

Using a real life scraped dataset we were not able to create a full ground truth; moreover we only had a partial knowledge of all the users in it. Considering this, we acted as an oracle and we tested our method on a small number of users that we knew for sure being accounts of the same person. \\
With only this topological information (up to now we had just used the neighbours similarity) we have found both bidirectional exact match and one-to-many matches. For the last case, we found also users that are not related, based on our knowledge. Varying the value of the signature size and the max distance, we're able to both find new correct match and to lose them. We suggest to find a trade-off between these values, according to the precision and the recall metrics.  \\
Unfortunately the downside we encountered using this feature was the discovery of a lot of false positives belonging to people that we know to be different from the main sockpuppet account we were looking for. \\

Network Science can provide a very effective framework to tackle the search of sockpuppet accounts in OSN, and it can be an interesting research direction because of the possibility of exploiting the structure of interactions that emerges with time. Furthermore this is something that is not OSN-dependent making it possible to hypothetically create an interaction graph from every OSN. \\  As future works we also plan to extend the experimentation involving labeled datasets and trying to combine other features mined from the graph. We have also started to explore the usage of lexical features that can be extracted from sent messages, using them in combination with the previously described ones, in order to investigate the influence of structural informations and linguistic ones. The first results are really promising and we are continuing to improve our research on the subject. 
\section*{Acknowledgements}
This work is supported by the scheme 'INFRAIA-01-2018-2019: Research and Innovation action', Grant Agreement n. 871042 'SoBigData++: European Integrated Infrastructure for Social Mining and Big Data Analytics'.

\bibliographystyle{splncs_srt}
\bibliography{biblio}

\begin{thebibliography}{1}

\bibitem{10.1145/276698.276876}
Indyk, P., Motwani, R.:
\newblock Approximate nearest neighbors: Towards removing the curse of
  dimensionality.
\newblock In: Proceedings of the Thirtieth Annual ACM Symposium on Theory of
  Computing. STOC '98, New York, NY, USA, Association for Computing Machinery
  (1998)  604–613

\bibitem{sckpgraph}
Li, J., Yuan, C., Zhou, W., Wang, J., Hu, S.:
\newblock Who are controlled by the same user? multiple identities deception
  detection via social interaction activity (student abstract).
\newblock Proceedings of the {AAAI} Conference on Artificial Intelligence
  \textbf{34}(10) (apr 2020)  13853--13854

\bibitem{10.1145/1242572.1242592}
Manku, G.S., Jain, A., Das~Sarma, A.:
\newblock Detecting near-duplicates for web crawling.
\newblock In: Proceedings of the 16th International Conference on World Wide
  Web. WWW '07, New York, NY, USA, Association for Computing Machinery (2007)
  141–150

\bibitem{DBLP:journals/corr/SolorioHM13}
Solorio, T., Hasan, R., Mizan, M.:
\newblock Sockpuppet detection in wikipedia: {A} corpus of real-world deceptive
  writing for linking identities.
\newblock CoRR \textbf{abs/1310.6772} (2013)

\bibitem{YAMAK2018124}
Yamak, Z., Saunier, J., Vercouter, L.:
\newblock Sockscatch: Automatic detection and grouping of sockpuppets in social
  media.
\newblock Knowledge-Based Systems \textbf{149} (2018)  124 -- 142

\end{thebibliography}

\end{document}